\documentclass[prl,twocolumn,showpacs,amsmath,amssymb]{revtex4}
\usepackage{graphicx}
\usepackage{texdraw}

\begin{document}

\title{Incommensurate Single-Angle Spiral Orderings of Classical
Heisenberg Spins on Zigzag Ladder Lattices}

\author{Yu.~I. Dublenych}
\affiliation{Institute for Condensed Matter Physics, National
Academy of Sciences of Ukraine, 1 Svientsitskii Street, 79011
Lviv, Ukraine}
\date{\today}
\pacs{05.50.+q, 75.10.Hk, 75.25.-j}

\begin{abstract}{Exact and rigorous solutions of the ground-state
problem for the classical Heisenberg model with nearest-neighbor
interactions on two- and three-dimensional lattices composed of
zigzag (triangular) ladders are obtained in a very simple way, with
the use of a cluster method. It is shown how the geometrical
frustration due to the presence of triangles as structural units
leads to the emergence of incommensurate spiral orderings and their
collinear limits. Interestingly, these orderings are determined by a
single angle (along with the signs of the interactions between
neighboring spins); therefore, the term ``single-angle spiral
orderings'' is proposed.}
\end{abstract}

\maketitle

In geometrically frustrated systems, the energies of all
interactions cannot be simultaneously minimized because of geometric
constraints \cite{bib1}. There is no general algorithm for
determining ground states even for classical frustrated systems. For
this reason, the ground-state problem for many classical Heisenberg
spin models still remains a challenging task for theorists
\cite{bib2,bib3}. The most known and used method for these purposes
is that of Luttinger and Tisza \cite{bib4} and its generalizations
\cite{bib5}. However, it appears not to be useful for complicated
Hamiltonians (for instance, with an external field or a biquadratic
exchange) and difficult to apply to non-Bravais lattices. There is
also a cluster method proposed by Lyons and Kaplan half a century
ago \cite{bib6}, but it is ``rather unknown,'' although simple and
intuitively clear \cite{bib7,bib8}. We developed this method in our
studies of the ground states for some Ising-type models
\cite{bib9,bib10,bib11}. Here, we consider the classical Heisenberg
model on a set of lattices for which, despite the presence of
frustration and a non-Bravais character of the lattices, a solution
of the ground-state problem can be obtained in a remarkably simple
and clear way, with the use of the cluster method. These are two-
and three-dimensional lattices composed of zigzag (triangular)
ladders. There are quite a few compounds where magnetic atoms are
arranged in lattices of this type
\cite{bib12,bib13,bib14,bib15,bib16,bib17,bib18,bib19,bib20,bib21,bib22}.
Making use of the cluster method, we clearly and rigorously show how
the presence of structural triangles and, hence, the geometric
frustration, leads to the emergence of incommensurate spiral
orderings. An important point is that, despite the presence of
several interaction parameters, these orderings are determined by a
single angle (together with the signs of the interactions between
neighboring spins along the ladder rungs). Therefore, we introduce
the term ``single-angle spiral orderings.''

Since triangular ladders are composed of identical isosceles
triangles, let us start with a single isosceles triangle and
classical spins (unit 3 vectors) at its vertexes (Fig.~1). $L$ and
$K$ are the strengths of couplings, and $\alpha$, $\beta$, and
$\gamma$ are the angles between the spins.

\begin{figure}[]
\begin{center}
\includegraphics[scale = 0.65]{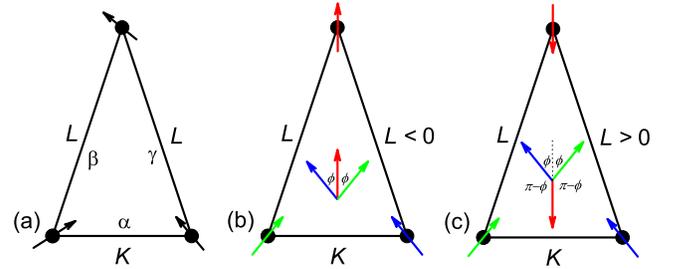}
\caption{(Color online) (a) An isosceles triangle with spins (unit 3
vectors) at its vertexes. $L$ and $K$ are the strengths of
couplings, and $\alpha$, $\beta$, and $\gamma$ are the angles
between the spins. In the $K < 0$ case, there is no frustration and
all the three spins are collinear. In the $K > 0$ case, there is a
frustration. At fixed value of $\alpha = 2\phi$, the minimum of the
energy is attained when two other angles are equal to $\phi$ if (b)
$L < 0$ or to $\pi - \phi$ if (c) $L > 0$. The spins are then
coplanar.}
\label{fig1}
\end{center}
\end{figure}

The energy of the system reads
\begin{equation}
E = K\cos\alpha + L(\cos\beta + \cos\gamma).
\label{eq1}
\end{equation}
What are the values of $\alpha$, $\beta$, and $\gamma$ that minimize
the energy? The $K < 0$ case is trivial: the spins are collinear,
because there is no frustration, that is the interaction energy can
be minimized by pairs. If $K > 0$ (the case of frustration), then,
at a fixed angle $\alpha = 2\phi$, the minimum value of the energy
is attained when $\cos\beta + \cos\gamma$ is maximum or minimum,
depending on the sign of $L$. Since $\alpha + \beta + \gamma
\leqslant 2\pi$, it is easy to see that, in the $L < 0$ case, the
sum is maximum, if the spins are coplanar and $\beta = \gamma =
\alpha/2 = \phi$ [Fig.~1(b)], and in the $L > 0$ case, this sum is
minimum, if the spins are coplanar and $\beta = \gamma = \pi -
\alpha/2 = \pi - \phi$ [Fig.~1(c)].

The ground-state energy of the triangle equals to
\begin{equation}
E_{gs} = K\cos2\phi - 2|L|\cos\phi.
\label{eq2}
\end{equation}
The energy is minimum if $\cos\phi = \frac{|L|}{2K}$. Thus, at $|L|
\leqslant 2K$, the ground state is determined by the angle $\phi$
and the spins are coplanar (the polarization plane, the orientation
of one of the three spins, and the chirality are arbitrary),
otherwise the spins are collinear.

Consider now a triangular ladder (Fig.~2). It is composed of
identical triangle clusters. Each $J$-bond belongs to two such
clusters; therefore, one should minimize the energy of a triangle
cluster with the interactions $J_0$ and $J/2$. In view of the
previous result, we can assert that the ground state is spiral and
determined by the angle $\phi = \arccos \frac{|J|}{4J_0}$, if $|J|
\leqslant 4J_0$, otherwise all the spins are collinear. Two examples
of the ground-state ordering for a triangular ladder are shown in
Fig.~2. It should be emphasized that all the spins are coplanar but
the polarization plane is arbitrary. Since the angle $\phi$ depends
on the strengths of couplings only, the spin ordering is, in
general, incommensurate along legs.

\begin{figure}[]
\begin{center}
\includegraphics[scale = 0.8]{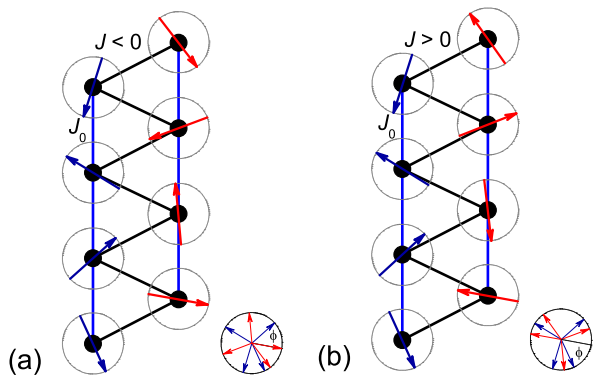}
\caption{(Color online) Ground-state configurations of a triangular
ladder with (a) $J < 0$ and (b) $J > 0$. $J_0 > 0$. The rotational
angle along legs is equal to $2\phi = 2\arccos\frac{|J|}{4J_0}$. All
the spins are coplanar. Here, the spins are in the plane of the
ladder but the polarization plane can be arbitrary, for instance,
perpendicular to the ladder. The orientation of one arbitrarily
chosen spin and the chirality (the same for both legs) can be
arbitrary as well.}
\label{fig2}
\end{center}
\end{figure}

It is of interest to consider a 2D-lattice composed of two types of
triangular ladders with nearest-neighbor interaction $J_0$ along
legs and $J_1$ and $J_2$ along rungs (Fig.~3). The lattice can be
partitioned into identical two-triangle clusters. Each $J_1$- and
$J_2$-bond belongs to two such clusters; therefore, one should
minimize the energy of a two-triangle cluster with the interactions
$J_0$, $J_1/2$, and $J_2/2$. The $J_0 < 0$ case is trivial: there is
no frustration. Consider the $J_0 > 0$ case. Let the angle between
the vectors at the ends of the $J_0$-bond be \emph{fixed} and equal
to $2\phi$ ($0 \leqslant \phi \leqslant \pi/2$). To minimize the
energy, the angles between neighboring spins joint by the
$J_i$-bonds ($i$ = 1, 2) should be equal to $\phi$ if $J_i < 0$ or
to $\pi - \phi$ if $J_i > 0$. The energy per cluster is
\begin{equation}
E_{gs} = J_0\cos2\phi - (|J_1| + |J_2|)\cos\phi.
\label{eq4}
\end{equation}
The angle $\phi$ that minimizes this energy is given by
\begin{equation}
\cos\phi = \frac{|J_1| + |J_2|}{4J_0}.
\label{eq5}
\end{equation}

\begin{figure}[]
\begin{center}
\includegraphics[scale = 0.8]{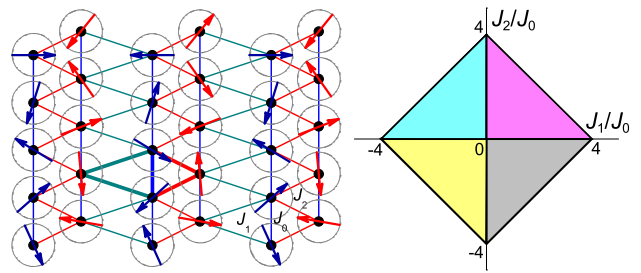}
\caption{(Color online) Left: A generalized triangular lattice and a
spin configuration on it for $J_1 < 0$ and $J_2 > 0$ ($J_0 > 0$,
$|J_1| + |J_2| \leqslant 4J_0$). A two-triangle cluster is shown by
bold lines. Right: Square of spiral phases for the generalized
triangular lattice. As the border of the square is approached, $\phi
= \arccos\frac{|J_1| + |J_2|}{4J_0}$ tends to zero and the spiral
phases continuously evolve to their collinear limits.}
\label{fig3}
\end{center}
\end{figure}

If $|J_1| + |J_2| \leqslant 4J_0$, then the spin ordering is spiral
and determined by the angle $\phi$ and by the signs of $J_1$ and
$J_2$, otherwise all the spins are collinear. In accordance with the
signs of $J_1$ and $J_2$, there are four spiral phases. The square
of spiral phases is shown in Fig.~3 (right panel). One of these are
depicted in Fig.~3 (left panel). As the border of the square is
approached, $\phi$ tends to zero and the spiral phases continuously
evolve to their collinear limits where the spins belonging to the
same leg have the same direction, although $J_0 > 0$. If $J_2 = J_1
= J_0 > 0$, then we have a well-known $120^\circ$-phase on the
triangular lattice.

Let us pass to a 3D zigzag ladder lattice, shown in Fig.~4. We refer
to it as a generalized hollandite lattice. The ground states of a
classical Heisenberg model on the hollandite lattice (dark cyan and
green rungs are identical) were studied in Ref.~\cite{bib21}. There
one can find an overview of magnetic compounds with this lattice
(see also Ref.~\cite{bib22}). Since the hollandite lattice is not a
Bravais one, a straightforward application of the Luttinger-Tisza
method is impossible; therefore, the authors used a modification of
it to find the ground-state energy; they also performed numerical
simulations. The cluster method makes it possible to solve this
ground-state problem in a very simple and clear way.

The generalized hollandite lattice is composed of three types of
zigzag ladders. Let the strengths of coupling between neighboring
spins along legs of all the ladders be $J_0$ (blue bonds) and $J_1$
(red bonds), $J_2$ (green bonds), and $J_3$ (dark cyan bonds) along
their rungs (Fig.~4). It is easy to see that this lattice can be
partitioned into identical three-triangle clusters shown in Fig.~4.

Each $J_1$-, $J_2$-, and $J_3$-bond belongs to two such clusters;
therefore, one should minimize the energy of a three-triangle
cluster with the interactions $J_0$, $J_1/2$, $J_2/2$, and $J_3/2$.
In the $J_0 < 0$ case, there is no frustration. Consider the $J_0 >
0$ case. Let the angle between the spins at the ends of the
$J_0$-bond be \emph{fixed} and equal to $2\phi$ ($0 \leqslant \phi
\leqslant \pi/2$). To minimize the energy, the angles between
neighboring spins joint by the $J_i$-bonds ($i$ = 1-3) should be
equal to $\phi$ if $J_i < 0$ or to $\pi - \phi$ if $J_i > 0$ (see
Fig.~1). The energy per cluster (or per site) equals to
\begin{equation}
E_{gs} = J_0\cos2\phi - (|J_1| + |J_2| + |J_3|)\cos\phi.
\label{eq8}
\end{equation}
The angle $\phi$ that minimizes the energy is given by
\begin{equation}
\cos\phi = \frac{|J_1| + |J_2| + |J_3|}{4J_0}.
\label{eq9}
\end{equation}

\begin{figure}[]
\begin{center}
\includegraphics[scale = 1.0]{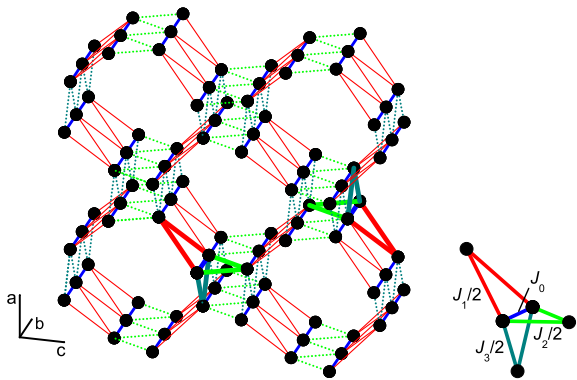}
\caption{(Color online) The generalized hollandite lattice is
composed of identical three-triangle clusters.}
\label{fig4}
\end{center}
\end{figure}

It is clear that all the spins on the lattice should be coplanar
that is parallel to a plane but this plane can be arbitrary. If the
plane, the orientation of one arbitrarily chosen spin, and the
chirality are fixed, then the distribution of angles completely
determines the ground-state spin configuration of the lattice at
fixed couplings.

Each set of signs of the interactions $J_1$, $J_2$, and $J_3$ ($J_0
> 0$) corresponds to a spiral phase and its collinear limit. Since
there are eight sets of signs, there are also eight spiral phases
and eight collinear limits. The octahedron of spiral phases and two
spiral orderings with their collinear limits are shown in Fig.~5.

\begin{figure}[]
\begin{center}
\includegraphics[scale = 1.0]{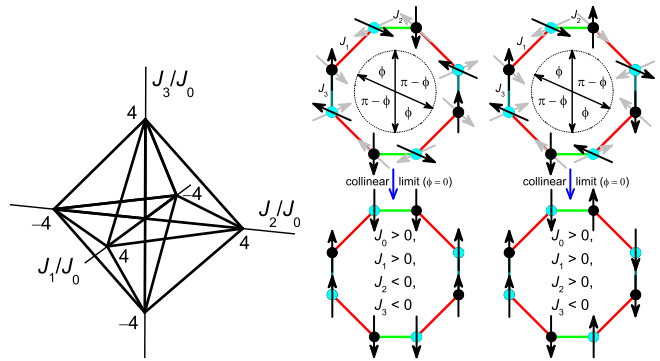}
\caption{(Color online) Left: Octahedron of spiral phases for the
classical Heisenberg model on the generalized hollandite lattice.
Right: Two examples of spiral orderings and their collinear limits.
Here, all the spins are parallel to the plane of the figure but this
plane can be arbitrary. Cyan and black circles denote sites over and
under the plane of the figure, respectively. Light gray arrows
represent spins next to ``black'' ones along legs.}
\label{fig5}
\end{center}
\end{figure}

The cluster shown in Fig.~4 gives the ground-state orderings of
classical Heisenberg spins on many other zigzag ladder lattices (see
Fig.~6). Eq.~(6) and the octahedron of spiral phases (Fig.~5) are
common for all these lattices. Hexagonal zigzag ladder lattices
similar to those shown in Figs.~6(a) and 6(b) have been identified
in many magnetic compounds
\cite{bib13,bib14,bib15,bib16,bib17,bib18,bib19,bib20}.

\begin{figure}[]
\begin{center}
\includegraphics[scale = 0.2]{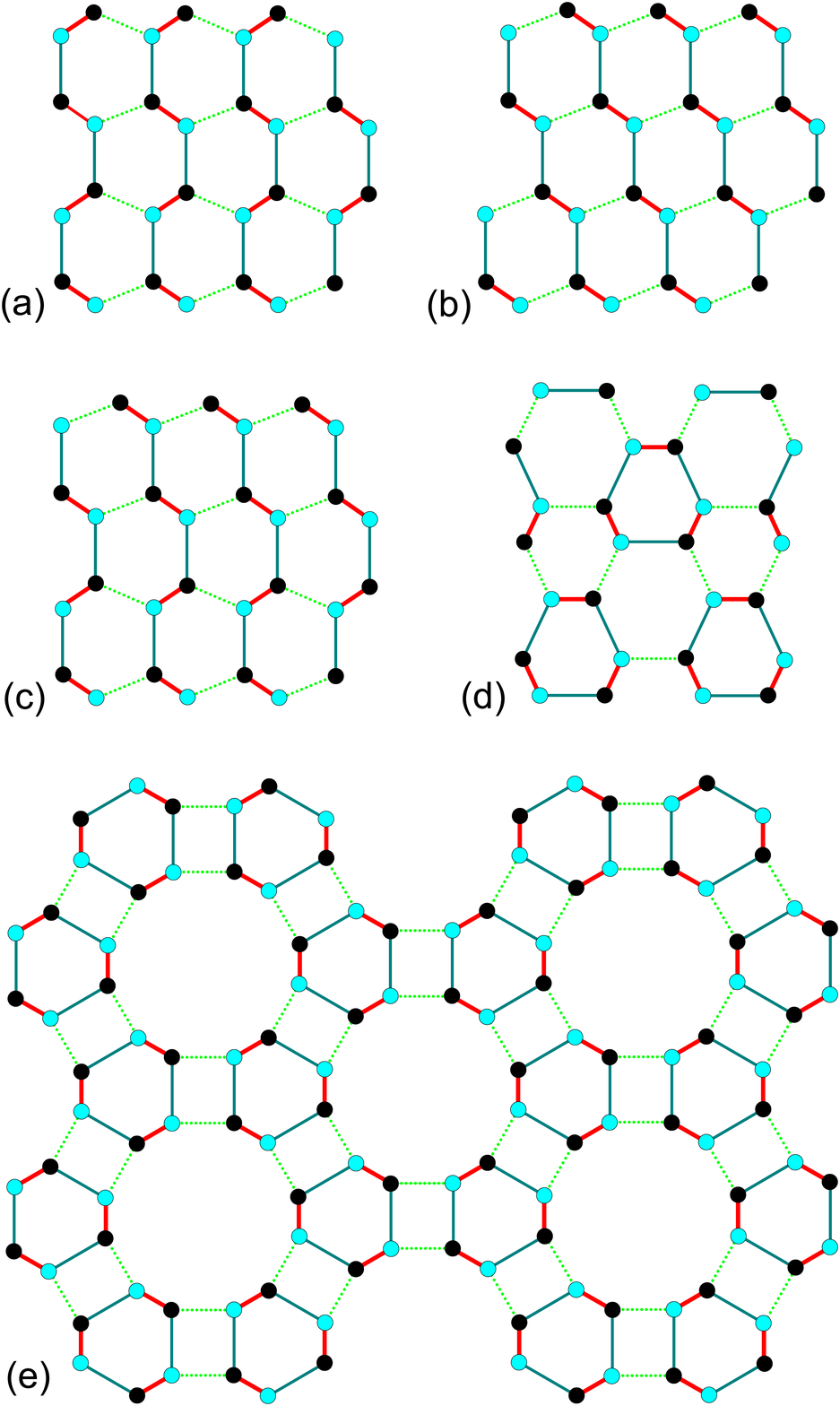}
\caption{(Color online) (a) - (d) Some hexagonal zigzag ladder
lattices. (e) A zigzag ladder lattice based on the truncated
trihexagonal tiling 4.6.12. The lattice (c) is a disordered mixture
of the lattices (a) and (b). All these lattices are composed of
identical clusters shown in Fig.~4. Cyan and black circles denote
sites over and under the plane of the figure, respectively. The legs
of zigzag ladders are perpendicular to the plane of the figure.}
\label{fig6}
\end{center}
\end{figure}

It is also interesting and instructive to consider a generalized
triangular lattice composed of three types of zigzag ladder (Fig.~7,
left panel). This lattice cannot be partitioned into identical
two-triangle clusters. Three types of two-triangle clusters have to
be considered. It can be shown that a single-angle spiral ordering
of classical Heisenberg spins is also possible in some regions of
the parameter space of this model. Let us distribute the energy of
each rung between two triangular plaquettes which share it in the
manner shown in Fig.~7 (left panel). Coefficients $\alpha_i$ for
this distribution can be arbitrary but should satisfy the conditions
$0 \leqslant \alpha_i \leqslant 1$.

\begin{figure}[]
\begin{center}
\includegraphics[scale = 1.00]{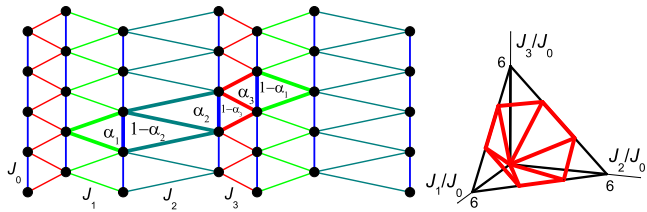}
\caption{(Color online) Left: A generalized triangular lattice,
composed of three types of zigzag ladders, and three types of
two-triangle clusters into which it can be partitioned. Right:
Hexagonal pyramid (in red) of a single-angle spiral phase on this
lattice for $J_i \geqslant 0$ ($i$ = 0-3).}
\label{fig7}
\end{center}
\end{figure}

The energies of two-triangle clusters can be written as
\begin{eqnarray}
&&E_{12} = J_0\cos2\phi_{12} - 2(\alpha_1 |J_1| + (1 - \alpha_2)|J_2|)\cos\phi_{12},\nonumber\\
&&E_{23} = J_0\cos2\phi_{23} - 2(\alpha_2 |J_2| + (1 - \alpha_3)|J_3|)\cos\phi_{23},\nonumber\\
&&E_{31} = J_0\cos2\phi_{31} - 2(\alpha_3 |J_3| + (1 - \alpha_1)|J_1|)\cos\phi_{31}.\nonumber\\
\label{eq10}
\end{eqnarray}
The condition for the minimum of these energies along with the
condition $\phi_{12} = \phi_{23} = \phi_{31} = \phi$ lead to the set
of equations
\begin{eqnarray}
\alpha_1 |J_1| + (1 - \alpha_2)|J_2| = \alpha_2 |J_2| + (1 - \alpha_3)|J_3| \nonumber\\
= \alpha_3 |J_3| + (1 - \alpha_1)|J_1| = \frac{|J_1| + |J_2| + |J_3|}{3}.
\label{eq10}
\end{eqnarray}
The angle $\phi$ that minimizes the energies is then given by
\begin{equation}
\cos\phi = \frac{|J_1| + |J_2| + |J_3|}{6J_0}.
\label{eq11}
\end{equation}

It it easy to show that the region in $(J_1/J_0, J_2/J_0,
J_3/J_0)$-space, where Eqs.~(8) and the inequalities $0 \leqslant
\alpha_i \leqslant 1$ are satisfied, is the polyhedral cone
determined by six vectors: (0, 2, 1), (0, 1, 2), (1, 0, 2), (2, 0,
1), (2, 1, 0), and (1, 2, 0) in the first octant (Fig.~7, right
panel) and similar polyhedral cones in other octants (not shown in
the figure). The single-angle spiral phases exist in the parts of
this cones which are bounded by the planes $|J_1| + |J_2| + |J_3| =
6J_0$. These are hexagonal pyramids, one of which is shown in Fig.~7
(right panel). Other parts of the polyhedral cones correspond to the
collinear limits of the spiral phases. It seems that, out of the
polyhedral cones, more complicated structures exist but here we
study single-angle orderings only. It is easy to extend these
results to arbitrary number of types of ladders.

\begin{figure}[]
\begin{center}
\includegraphics[scale = 0.23]{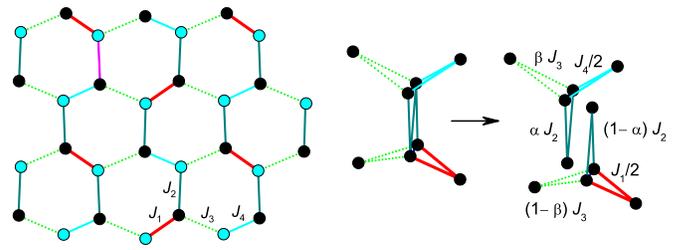}
\caption{(Color online) A hexagonal zigzag ladder lattice realized
in the $S = \frac32$ compound $\beta$-CaCr$_2$O$_4$ and its
partition into two types of three-triangle clusters. Cyan and black
circles denote sites over and under the plane of the figure,
respectively. The legs of zigzag ladders are perpendicular to the
plane of the figure.}
\label{fig8}
\end{center}
\end{figure}

Finally, consider a  hexagonal zigzag ladder lattice shown in
Fig.~8. This lattice is realized in the $S = \frac32$ compound
$\beta$-CaCr$_2$O$_4$ \cite{bib18,bib19} (however, the spin ordering
observed in this compound is more complicated than obtained here).
In contrast to the lattices shown in Figs.~6(a)-6(d), this lattice
can be partitioned only into two types of three-triangle clusters
(Fig.~8, right panel). The ground-state energies for these clusters
are given by
\begin{eqnarray}
&&E_1 = J_0\cos2\phi_1 - (2\alpha |J_2| + 2\beta |J_3| + |J_4|)\cos\phi_1, \nonumber\\
&&E_2 = J_0\cos2\phi_2\nonumber\\
&&~~- \left[|J_1| + 2(1 - \alpha)|J_2| + 2(1 - \beta)|J_3|\right]\cos\phi_2.
\label{eq12}
\end{eqnarray}
Minimizing these energies with respect to $\phi_1 = \phi_2 = \phi$,
we obtain
\begin{equation}
\cos\phi = \frac{|J_1| + 2|J_2| + 2|J_3| + |J_4|}{8J_0}.
\label{eq13}
\end{equation}

The conditions $0 \leqslant \alpha \leqslant 1$ and $0 \leqslant
\beta \leqslant 1$ lead to the inequality
\begin{equation}
||J_4| - |J_1|| \leqslant 2(|J_2| + |J_3|).
\label{eq14}
\end{equation}
This inequality along with the inequality $|J_1| + 2|J_2| + 2|J_3| +
|J_4| \leqslant 8J_0$ determine the regions for existence of
single-angle spiral phases.

To summarize, we have rigorously proven that, in systems of
classical Heisenberg spins on zigzag (triangular) ladder lattices,
there exist incommensurate ground-state spiral orderings (phases)
characterized by a single angle and by the signs of the interactions
between neighboring spins (along ladder rungs). We propose to name
these orderings (phases) ``single-angle spiral orderings (phases).''

\end{document}